# UNDERSTANDING THE ELUSIVE DWARF CARBON STAR

*Ender Bulunan Cüce Karbon Yıldızını Anlama*

PAUL J. GREEN

*Smithsonian Astrophysical Observatory*
*Harvard-Smithsonian Center for Astrophysics*
*60 Garden St., Cambridge, MA 02140, USA*

**Abstract.** Most stars in our Galaxy with photospheric C/O > 1 (carbon stars) are not giants but dwarfs. The newly-recognized class of dwarf carbon stars joins the growing family of stars with peculiar abundances that are now recognized as products of mass-transfer binary (MTB) evolution. The dozen examples now known span a wide range of evolutionary histories, ages, and abundances. These stars can already provide some much-needed constraints on the formation of AGB C stars in the disk and spheroid populations, and on the parameters characterizing binary evolution there. A larger sample, with some bright members, would hasten our progress.

## 1. Introduction

The 'C' and the star on the Turkish flag make this conference a good place to set the stage for a new perspective on carbon stars. Just so we get it straight, the cool, luminous AGB stars that dominate C-star theory and observation (and the rest of this conference) are in all likelihood atypical. If a carbon star is defined as a star showing carbon molecular features ($C_2$ bands, with CH, CN and/or s-process enhancements often associated), then the vast majority of C stars are not giants at all but dwarfs. Recognition of this paradigm shift is at first glance troublesome, both for observers, who can no longer assume that a C-star spectrum implies a giant luminosity, and for theorists, who must explain how a dwarf can show C/O > 1 when no carbon is produced by main-sequence hydrogen burning. I'll explain briefly how we came to recognize that dwarf carbon (dC) stars are not some rare freak of nature, but constitute a whole class of stars that fall naturally into the family of post mass-transfer binary systems.

## 2. Finding Faint High-Latitude Carbon Stars

Our initial search for Faint High Latitude Carbon (FHLC) stars was motivated by the impression that models of the chemical and dynamical properties of the Galactic spheroid (the 'halo') are still rather weakly constrained. In the grand scheme, did a monolithic protogalaxy undergo rapid collapse and enrichment (Eggen, Lynden-Bell & Sandage 1962), or did many smaller dwarf galaxies merge together (Searle & Zinn 1978)? Both processes may contribute, with mergings ongoing today. More modestly, how can we best characterize the mass-to-light ratio, the velocity ellipsoid and systemic rotation of the outer halo? Intrinsically bright stars visible to large galactocentric distances ($10-100$ kpc) provide the best opportunity. Faint C stars have been sought as excellent tracers of the outer halo because they were thought to be distant giants, and because they are readily recognizable from their strong $C_2$ and CN absorption bands. As tracers, they do not suffer as do globular clusters from selection effects complicated by tidal interactions with the disk.

Objective-prism photography with wide-field Schmidt telescopes has yielded low-dispersion spectra for thousands of objects over substantial portions of the sky. At high galactic latitudes, we find mostly CH stars, and possibly some R stars. (Unless stated explicitly, it is to these warmer types that I refer here.) Fewer than 1% of the 6000 stars in Stephenson's (1989) catalogue are the faint, high-latitude carbon (FHLC) stars ($V > 13$, $|b| > 40°$) most useful as dynamical probes of the outer halo. The two most prolific sources of published FHLC stars, the Case low-dispersion survey (CLS; Sanduleak & Pesch 1988) and the University of Michigan – Cerro Tololo survey (UM; MacAlpine & Williams 1981) appear to be complete to about $V = 16$ and have provided about 30 FHLC stars. Emission-line objects, not FHLC stars, were the primary goal of these photographic surveys, and known FHLC stars were not examined to help predefine selection criteria or estimate completeness. The surface density of FHLC stars from objective-prism surveys is low, about one per 50 deg$^2$ to $V \sim 16$. The challenge was then to go deeper than the objective-prism surveys, while still covering a wide area with CCD imaging. We sought to expand shallower objective-prism survey samples to provide a significant-sized sample of distant halo C stars. The combined sample could be used for tests of the spheroid density law, and for dynamical analysis. We used a 3-filter, two-color photometric technique to distinguish C stars from other late-type stars via intermediate band ($\approx 200$ Å FWHM) filters (e.g. Cook & Aaronson 1989; Palmer & Wing 1982). One ("77") is centered on a region of TiO absorption near $\lambda 7752$ Å, and the other ("81") is on a CN absorption band near $\lambda 8104$ Å. The $77-81$ color thus separates C stars from other stars of similar effective temperature (and similar $V - I$) because C stars appear particularly faint in the 81 filter relative to the 77 filter, while the converse prevails in M stars. By the end, our CCD survey covered 52 deg$^2$ of high galactic latitude sky to a depth of about $V = 18$. Only one highly ranked $V = 17$ candidate was found to have strong carbon and CN bands. That star, ($\alpha_{1950} = 03^h11^m44^s.08$, $\delta = +07°33'38.0''$), is of course my favorite in the whole sky. So perhaps the hubris of using one star to estimate a surface density will be pardoned. To a depth of $V = 18$, the surface density of FHLC stars in our CCD survey is about 0.02 deg$^{-2}$ (Green et al. 1994), the same as the surface density from objective-prism surveys. We went deeper, so why didn't we find more? First, there might well be more FHLC stars to $V = 18$ than to $V = 16$, but we can't tell within the large uncertainties. If the apparent difference is real, it could be that we begin to "run out of Galaxy" at these magnitudes. Also remember that most photographic surveys pick up Swan bands of $C_2$, while our survey picks up CN, so that slightly different trends or normalization may obtain with metallicity and/or galactocentric radius.

## 3. Stumbling Over Dwarfs

While taking follow-up optical spectra of candidate FHLC stars from our survey at the KPNO 2.1-m telescope, we also obtained numerous spectra of previously known FHLC stars (Green & Margon 1990). One of these objects, CLS 96, also bore the comment "LP 328-57?" in the Sanduleak & Pesch (1988) list, indicating a possible identification with a high-proper-motion (p.m.) star. Bidelman and MacConnell had independently noticed this association of CLS 96 with the Luyten p.m. star at around the same time. When Peter Pesch alerted us to this, we immediately checked the p.m. catalogues (via SIMBAD) for other faint C stars and found another probable association, that of LHS 1075 and C*22 from the UM survey. Astrometry using the 30-year baseline between the POSS and the HST Quick V Survey plates yielded yet another high p.m. C star, CLS 31. Large proper motions could only occur at small distances for a star bound to the Galaxy, thus mandating that these faint C stars must be dwarfs.

Up until then, the only known flaw in the otherwise promising technique of using FHLC stars as halo tracers had been the existence of one lone star, G77–61, a $V = 13.9$ dC also with a high proper motion (Dahn et al. 1977). The object was subsequently shown to be a single-lined spectroscopic binary of period 245 d and parallax $\pi_{abs} = 0.017 \pm 0.003''$ (Dearborn et al. 1986). These authors argued that, since main-sequence stars can't dredge up carbon, the most reasonable explanation for the prominent $C_2$ bands in the dwarf's spectrum is photospheric deposition of mass from a now unseen companion during the companion's second ascent of the giant branch. Even given such different origins, we found that the optical spectra of dCs at resolution $\geq 1$ Å are strikingly similar to those of Pop II carbon giants, from their wide distribution of $C^{13}/C^{12}$ ratios to their enhanced $s$-process abundances (Green & Margon 1994). 'Faint' doesn't guarantee 'distant' anymore; C stars are nearly on the same footing as other late-type stars.

How could we know if there remained dCs of lower p.m. in the FHLC star sample? Since IR colors are commonly used to determine C giant luminosities, I took the time to plot the published IR colors of FHLC stars (e.g. Bothun et al. 1991; Mould et al. 1985). The three dwarfs were redder in $H - K$ than most other C stars, but so were two or three other stars. A more thorough p.m. survey of all known FHLC stars (Green et al. 1992) showed that the other odd-colored stars were also moving! This brought the total up to five and strongly suggested that dCs may have $JHK$ colors distinct enough to identify them as dwarfs.

## 4. Reign of the Dwarfs

Our Monte Carlo simulations indicated that the proper motion survey, sensitive to p.m. $> 0.1''$/year, could detect Pop II dCs brighter than $V = 18$ 98% of the time. Therefore, since we detected proper motions for 5 of 39 FHLC stars in our survey, we know that at least 13% of FHLC stars to that magnitude are dCs. Presumably, deeper surveys will find higher fractions, because they will begin to probe beyond the Galaxy for giants. But as it stands, how does the space density of dCs locally compare with other types of C stars? Simply taking 13% of FHLC stars, with their a surface density of 0.02 deg$^{-1}$, yields about 100 dCs. All dCs with absolute magnitude estimates to date have $M_V \sim 10$. Assuming this holds for all dCs, a survey limit of $V = 18$ corresponds to a sphere of 400 pc radius. Within this volume, the number of dCs easily surpasses the sum of *all other types of carbon stars combined*, including N, R, CH, Ba and sgCH (or dBa) stars (Green et al. 1992). My simplistic calculation is nevertheless quite conservative since all these latter types will be brighter by several magnitudes than dCs within the same volume. Furthermore the estimated C dwarf to C giant fraction must be a lower limit, because it concerns only high-proper-motion dCs. The binary mass transfer explanation for photospheric carbon in dCs predicts that they should exist in the disk as well. Disk dCs would tend to have small proper motions. If, for example, 2 disk dCs in our sample were counted incorrectly as giants, then the true fraction is closer to 20%.

Soon after our p.m. survey, Warren et al. (1992) found two faint dCs in the south by means of their high proper motions. Heber et al. (1993) found a composite spectrum DA/dC binary system, PG0824+289, with a 60,000 K white dwarf. The prototype dC G77–61 was also known to have a faint companion ($T_{\rm eff} < 6000$ K from an IUE upper limit), but PG0824+289 is truly the smoking gun, white hot evidence for the mass-transfer hypothesis. In addition, since PG0824+289 had no detectable proper motion and disk kinematics, it may represent the first known disk dC. Only a year passed before a similar DA/dC composite, CBS 311, was found (Liebert et al. 1994). Neither could have been found via proper motion selection. How many such dCs are out there?

## 5. Predicting the Space Density of Dwarf Carbon Stars

If the space density of disk dCs scales with the ratio of disk/halo space densities (at least $\sim 500$: Bahcall & Soneira 1984; Morrison 1993), then we would expect there to be about 1.3 deg$^{-2}$ to our nominal limiting magnitude of $V = 18$. But this exceeds the TOTAL surface density of FHLC stars by about a factor of 6. This back-of-the-napkin calculation shows that the dC fraction must actually be much lower in the disk than in the halo.

Age and metallicity differences offer a likely explanation for the larger fraction of dwarfs that are dCs in the spheroid compared to the disk. Other effects such as the means of accretion (e.g. Roche lobe overflow, or wind) and mixing efficiency in the accreting star currently preclude simple analytic predictions. De Kool & Green (1995) have constructed simulated samples of dC stars to determine whether reasonable assumptions lead to dC space densities compatible with observations, and to investigate how these assumptions affect the expected properties of dCs. A simulated population of dCs is constructed by following the evolution of a large number of binaries using simple analytic fits to detailed evolutionary calculations, and determining which ones would presently contain a dC star. The zero-age parameters of the sample are chosen randomly from observed distributions of unevolved binaries. The space density of halo dC stars that we predict ($\sim 2$–$4 \times 10^{-7}$ pc$^{-3}$) is in agreement with current observational constraints. The predicted local space density of disk dC stars ($\sim 1 \times 10^{-6}$ pc$^{-3}$) may be a bit high, since it still predicts nearly as many disk dCs as there are FHLC stars observed. The fraction of binaries that produces dCs depends strongly on initial metallicity, and virtually no dCs are formed in systems with an initial metallicity of more than half solar. Thus all disk dCs are predicted to be in binaries that formed in the very early phases of disk star formation, and their number depends strongly on assumptions about the age-metallicity relation during this epoch. The predictions for the halo are much less model-dependent. In either population, we may expect that dCs on average will exhibit lower than average metallicities. We also predict that dCs in the disk, for instance, will eventually be shown to have a scale-height consistent with old or thick-disk ages. The simulated dC orbital period distributions are bimodal, with one peak between $10^3$ and $10^5$ days and another peak between $10^2$ and $10^3$ days. The shorter-period component is caused by systems that have gone through a common envelope phase, while most have accreted from an AGB wind. The simulated period distributions bear a strong resemblance to the observed orbital period distribution of barium and CH giants, which may be the evolved descendants of the disk and halo dC populations we modeled.

## 6. Joining the Mass Transfer Binary (MTB) Family

Because they are considerably more luminous, barium (Ba), CH, and S stars have been better studied and characterized than dCs. These giants show peculiar abundances, with $C/O \geq 1$ and a strong overabundance of $s$-process elements, thought to be produced during shell burning on the AGB. Some S stars are indeed on the AGB. Those suspected of being MTB products (the 'extrinsic' S stars) show no sign of the unstable element technetium (Tc) in their spectra, unlike their AGB (or 'intrinsic') S analogs. Ba, CH and extrinsic (non-Tc) S stars are red giants that have not undergone the thermal AGB pulsations necessary to produce their observed peculiar photospheric abundances. Observations are consistent with the MTB hypothesis as an explanation for all of them. McClure & Woodsworth (1990) present orbital periods for CH stars; Jorissen & Boffin (1992) have collected orbital parameters and abundances for Ba stars, and Jorissen & Mayor (1992) for S stars. The mass functions derived for systems with detected periods are consistent with white dwarf companions. A fair fraction have no period yet detected, which almost certainly means that it is very long. This implies wind accretion, as does the non-zero eccentricity of many orbits. Diagrams of $e$ vs. $\log P$ can reveal the typical Roche lobe radius of the former AGB primaries in a sample, since tidal effects or Roche lobe overflow will circularize an orbit with separation near that radius.

CH stars are halo giants whose unevolved precursors could be halo dCs. Similarly, the Ba giants, with old-disk kinematics and near-solar metallicities, are likely to represent the high-mass end of the population of disk stars that have experienced mass transfer from an AGB companion. The CH subgiant stars (Smith, Coleman & Lambert 1993), maybe better described as Ba dwarfs (dBa), are main-sequence (MS) counterparts, or perhaps precursors of Ba giants (North, Jorissen & Mayor, this conference). Extrinsic S stars probably represent somewhat cooler, lower mass MTB disk giants (Jorissen & Mayor 1992) whose precursors have yet to be postulated. The spectrum of any post-MTB giant (CH, Ba, or extrinsic S star) may differ substantially from that of its unevolved dwarf precursor, particularly if the mass of the convective zone changes greatly during evolution.

Lower limits to ages of white dwarf companions (e.g. Johnson et al. 1993; Smith et al. 1993; Bohm-Vitense et al. 1984) generally exceed the lifetime of the giant phase, so that the mass transfer episode *must* have occurred while the contemporary giant was still on the main sequence. Since MS lifetimes are much longer than giant lifetimes, the MS precursors to Ba, CH, and extrinsic S giants should abound, but are clearly more difficult to recognize or detect. We may explain the unique existence of dCs in both disk and halo as a consequence of their mass range, the lowest of any post-mass-transfer objects so far discussed (mostly near $0.5 M_\odot$). Still, the undetected companion of G77–61 in contrast to the very hot DA companions to the dCs PG0824+289 and CBS 311 reveals a truly wide range of ages since mass transfer among dCs.

## 7. Finding More about dC Stars

I'll just summarize four areas that can be clarified by observations in the near future: (1) Long-term radial velocity monitoring is needed to prove duplicity, and to determine the mass function and eccentricity. These must be correlated with measured abundances. A trend predicted in our simulations, and observed in Ba giants, is a small but significant anti-correlation between *s*-process overabundance and orbital period. (2) Good abundance determinations are needed, particularly of Tc, heavy to light *s*-process element ratios, metallicity, C/O and carbon isotope ratios. More rapid progress in our understanding will also be made when model atmospheres for dCs have been matched to UV/optical/IR spectrophotometry and to luminosities derived from trigonometric parallax. (3) Does enriched accreted material mix beyond the shallow convective zones of higher-mass dwarfs? Proffitt & Michaud (1989) argued that the higher mean molecular weight of the accreted material leads to instability and mixing even into radiative layers. If there is no such mixing, the predicted number of disk dCs increases by an order of magnitude, and extends to higher-mass dwarfs. (4) More FHLC and more dC stars must be found. There may still be low p.m. dCs lurking even in the sample of currently known FHLC stars. We need to know whether $JHK$ colors are a good luminosity discriminant for disk dCs, and *why* they are a good discriminant for halo dCs. Ancient generations of AGB stars, the physics of mass transfer, and important parameters describing binary mass ratios and separations may be probed by such measurements.

Some of these suggestions for dCs mimic hard work already done on peculiar red giants, which have led to enormous leaps in our understanding of binary evolution. I hate to point out that the work on dCs may be even harder, because to date all are fainter than about 14th visual magnitude.

There are few surveys specifically designed to net new C stars, but we now know that deeper surveys should reveal more dCs locally. As an example, I have initiated a deep CCD multicolor Schmidt survey, and a survey using the CCD grism transit scans of Schmidt, Schneider & Gunn (1995). Multicolor searches for CH and dC stars in globular clusters have also begun, from which relative distance and age uncertainties are largely removed. We hope that the Sloan Digital Sky Survey will be a major source of new discoveries of FHLC and dC stars in the field. A large, well-quantified sample of dC stars will go a long way toward a better understanding of how evolution in mass transfer binary systems is affected by age, metallicity and other factors. Let's just say there's a lot of physics packed into these dwarfs, at once the most elusive and most common type of carbon star in the Galaxy.